\documentclass[12pt,a4paper]{article}
\usepackage[cp866]{inputenc}
\usepackage[T2A]{fontenc}
\usepackage[dvips]{graphicx}
\begin{document}
\def\r{g/cm${}^2$}
\def\d{\displaystyle}
\centerline {\bf Gravitational Mesolensing by King Objects}

\centerline {\bf and Quasar-Galaxy Associations }
\vskip 0.5cm
\centerline {Yu.V.Baryshev and Yu.L.Bukhmastova }
\vskip0.5cm
\centerline {\it Astronomical Institute, St. Petersburg State University,}
\centerline {\it St.Petersburg, Russia, bukh\_julia@mail.ru}
\vskip 0.7cm
{\bf 1. INTRODUCTION}
\vskip 0.3 cm
The question of whether there is a physical relation between quasars and bright 
galaxies has been discussed in the literature for 30 years [1-3; 4 ch.12].
In one recent work devoted to this problem [5], an automated search for close 
guasar-galaxy pairs was undertaken with angular resolution better than $10^{\prime}.$ 
The observed number of such pairs exceeds the number of expected random coincidences 
by a significant factor. For examaple, a simple estimate of the probability of finding 
a randomly-positioned object at an angular distance of $2^{\prime}$ from a galaxy with 
magnitude brighter than $15^m$ gives value of about  $10^{-3}.$  Consequently, the 
expected number of random pairs in a sample containing 5000 quasars is 5, while the 
catalog referred to above contains 38 such pairs. The possibility of a physical relation 
between quasars and galaxies in associations is also supported by the "angular separation-redshift" 
diagram for the galaxies, constructed for 392 pairs. According to this observational relationship, 
the mean linear separation between  the galaxy and quasar projected onto the celestial sphere
remains constant and  is equal  to about 100 kpc. Based on these data, Burbidge et al. [5]  
conclude  that quasars tend to reside  near the halos of normal galaxies far more  often 
than is expected for random projections, and that this physical relation demands an explanation. 
At least two different interpretations of this observed "Arp effect" are possible:

(1) Quasars in associations are located in the halos of nearby galaxies, so that their redshifts
are not cosmological, i.e., they do not reflect the distance to these objects.

(2) Quasars in associations are only projected onto the halos of nearby galaxies; their redshifts 
preserve their cosmological origin, and the increased apparent luminosity of such quasars is the 
result of gravitational lensing by objects located in the galactic halos.

In his papers [1-3], Arp holds to the first possibility, because no other compelling explanations
were proposed. Accordingly, he considers it necessary to accept the noncosmological nature of the
quasar redshifts, and to invoke new physics to interpret the observed effect [2].

Interpretation of quasar-galaxy associations as the result of gravitational lensing by galaxy halo
stars (microlensing) was first suggested in [6]. However, it was later shown [7,8,4] that the 
gravitational microlensing hypothesis cannot account for the observed effect because of the extremely 
small observed surface density of faint quasars that provide the sources for the microlensing. This gave
indirect support to the first interpretation, since it was thought that gravitational lensing in general
could not explain the occurrence of quasar-galaxy associations [4]. For example, in [3], it was concluded
that the existence of quasar-galaxy associations could, in principle, be explained using standard physics
if gravitational lensing played a significant role in the halos of bright galaxies. In particular, quasars
located in the dwarf satellites of bright galaxies could be explained by simply postulating the existence
of compact lensing objects at the ends of bridges connecting these satellites to the parent galaxy
(the ''lorgnette'' postulate, in Arp's terminology). However, Arp believes that several arguments strongly
oppose the gravitational micro-lensing hypothesis:

(1) In pairs with nearby galaxies, the quasars predominantly lie at distances of a few times the diameters
of the galaxy itself. In this case, the microlenses must occupy a huge space around their galaxies, and an
extremely high halo mass is required to account for the observed number of associations.

(2) Quasar-galaxy associations in galaxy groups are most often found not for the main galaxy, but for a 
companion galaxy in the group.

(3) Quasars with the highest redshift $z$ are associated with galaxies with the lowest $z.$

(4) Counts of faint quasars showed that their number is insufficient to provide the observed effect. It was 
concluded in [8, 4] that the quasar luminosity function has a gentler slope than required for strong gravitational
lensing.

(5) Analysis of archival data on quasar variability in associations [9] showed the absence of such variability on
timescales up to a few decades, which contradicts the microlensing hypothesis.

However, we will show here that these arguments do not, in general, contradict he gravi-tational microlensing hypothesis. 
In particular, if the gravitational lenses are not stars, but objects such as globular clusters, dwarf galaxies, or
clusterized hidden mass, all of the above problems can be removed. In addition, a quantitative interpretation of the
Arp effect based on the gravi-tational lensing hypothesis requires that distant nuclei of active galaxies be taken as 
the sources, rather than quasars, and that the nonuniform distribution of lenses along the line of sight be taken into
account. In Section 2, we describe the initial assumptions for the gravitational lensing model adopted. The basic
equations and results of calculations for extended lenses with King mass distributions are presented in Section 3.
We interpret the observed properties of quasar-galaxy associations in Section 4. Observational tests of the adopted
model are discussed in Section 5. Finally, in Section 6, we summarize our main conclusions.
\vskip 0.7cm

{\bf 2. INITIAL ASSUMPTIONS OF THE MODEL}
\vskip 0.4cm

In order to construct a gravitational lensing model, we must determine

(1) the lensed objects (sources);

(2) the lensing objects (lenses); 

(3) the distribution of lenses along the line of sight.

 \vskip 0.5cm
{\it 2.1. Sourses}
\vskip 0.5cm

The lensed sources are usually taken to be weak quasars. It follows from argument (4) in Section 1, however, that the
number of faint quasars is too small to provide a statistically significant effect. Here, we adopt the hypothesis of Barnothy 
and Tyson [10,11] that quasars, at least in part, are gravitationally amplified images of the nuclei of active galaxies. 
We accordingly adopt as the lensed sources compact massive objects located in the central regions of active galaxies, such
as Seyfert galaxies and radio galaxies of all types. The number of these objects is large enough to account for the observed 
of associations with reasonable brightness amplification coefficients. According to the latest observations made with
the Hubble space telescope, the surface density of normal galaxies with magnitudes $28^m-29^m$  is of the order of $10^6$gal./deg${}^2,$
which supports the result obtained by Tyson [12]. If  $1\%$ of such galaxies have active nuclei, we estimate that the lensed sources
have a surface density of  $10^4$ sources/deg${}^2$  at redshift $z_s\sim 1.$
\vskip 0.7cm

{\it 2.2. Lenses}
\vskip 0.5cm 

Papers concerning statistical analyses of gravitational lensing usually take the gravitational lenses to be point sources or 
isothermal spheres. Such models make it possible to obtain some analytic results, but cannot remove the contradictions associated
with the arguments presented above. We will consider the gravitational lenses to be objects with a King mass distribution, which is
widely encountered in nature and approximates well the density profiles of globular clusters, dwarf galaxies, and possibly other
self-gravitating objects with hidden mass.

Globular clusters were first considered as gravitational lens candidates in [13], in connection with the association between
the quasar 3C455 and the nearby galaxy NGC 7413, using a M15-type globular cluster as an example. In [14], Baryshev et al. proposed
globular clusters and other objects with mass $10^3\div 10^9 M_{\odot}$ as gravitational lenses ("mesolenses") in order to account
for the properties of quasar-galaxy associations. Here, we use the physical parameters of globular clusters given in [15] to obtain
specific quantitative results. We have the required data on core radius, total radius, central volume density, and mass for 56 globular
clusters (see the table 1).

 The radial dependence of the volume density of globular clusters is given by a King law [16],
 $$ \rho(p)=\cases{\rho_0\cdot(1+p^2/ r_c^2)^{-3/2},
&for $ p<r_t$;\cr
0,&for $p\geq r_t$\cr},\eqno(1)$$
where $p$ is the impact parameter, $\rho_0$ is thecentral volume density,
 $r_c$ is the core radius,and $r_t$ is the outer radius of the cluster.  When $r_t\gg r_c$, the  central surface density  $\sigma_0$[\r]       
can be  expressed in terms of the central volume density 
$$\sigma_0=2\rho_0 r_c \eqno(2) $$
or in terms  of the globular cluster mass :
$$\sigma_0= 10\Bigl(\frac{M}{10^6M_\odot}\Bigr)\Bigl(\frac{1\hbox{ pc}}{r_c}\Bigr)^2 \hbox{\r},\eqno(2a)$$
For the globular cluster sample [15], the ratio $r_t/r_c$ lies in the range  $5-630,$ while  $\sigma_0$  lies between $0.02$ and $140$ \r.  
The  surface density distribution of globular  clusters  is well represented by the logarithmic  Gaussian dependence 
$$f_{\sigma_0}(x)=\frac{1}{x\sigma\sqrt 2\pi}e^{-\d{\frac{{(\ln x-\mu)}^2}
{2\sigma^2}}} \eqno(3)$$
where $x=0.1\sigma_0, \hskip10pt\mu=0.1,\hskip10pt \sigma=3.$

The focal distance of a King lens is related to  $\sigma_0$ by  the expression [17]: 
$$F_L=3.5\Bigl(\frac{100\hbox{\r}}{\sigma_0}\Bigr)\hbox{ Mpc} \eqno(4)$$
The number of globular clusters in galactic  halos may constitute  from several  hundreds to thousands. Thus, globular clusters can be considere
good gravi-tational lens candidates for accounting for the observed quasar-galaxy associations. Globular clusters can manifest themselves at
distances of several Mpc from the observer or the source, corresponding to $\Delta z\geq 4*10^{-4}.$
\vskip 0.7cm
Table 1. Data on globular clusters
\vskip 0.3 cm
\vbox{\offinterlineskip
\hrule
\halign{&\vrule#& \strut\quad\hfil#\quad\cr
height2pt&\omit&&\omit&&\omit&&\omit&&\omit&&\omit&&\omit&\cr
&NGC\hfil& &${\lg\frac{ M}{M_{\odot}}}$\hfil& &$r_c$\hfil& &$r_t$\hfil& &$r_t/r_c$\hfil& &$\lg\rho_0 $\hfil& &$\sigma_0$&\cr
&\omit\hfil&&\omit\hfil&& pc\hfil&&pc\hfil&&\omit\hfil&&$M_\odot pc^{-3}$\hfil&&g cm${}^{-2}$&\cr
height2pt&\omit&&\omit&&\omit&&\omit&&\omit&&\omit&&\omit&\cr
\noalign{\hrule}
height2pt&\omit&&\omit&&\omit&&\omit&&\omit&&\omit&&\omit&\cr
&104&&6.1&&0.50&&79.24&&158.48&&5.1&&25.8&\cr
&288&&4.9&&3.46&&27.48&&7.94&&2.1&&0.18&\cr
&362&&5.5&&0.42&&58.87&&140.16&&4.7&&8.5&\cr
&1851&&6.0&&0.27&&85.38&&316.22&&5.7&&56.9&\cr
&1904&&5.2&&0.60&&37.85&&63.09&&4.2&&4.0&\cr
&2419&&5.6&&8.51&&213.76&&25.11&&1.4&&0.08&\cr
&2808&&6.2&&0.70&&55.6&&79.43&&4.9&&23.4&\cr
&3201&&5.5&&2.08&&41.50&&19.95&&3.0&&0.87&\cr
&4147&&4.7&&0.54&&42.89&&79.43&&3.7&&1.13&\cr
&4590&&5.0&&1.86&&93.22&&50.11&&2.5&&0.24&\cr
&5053&&4.7&&10.47&&66.06&&6.30&&0.6&&0.02&\cr
&5139&&6.6&&3.71&&58.79&&15.84&&3.5&&4.93&\cr
&5272&&5.8&&1.47&&147&&100&&3.5&&1.95&\cr
&5286&&5.5&&0.77&&24.34&&31.62&&4.3&&6.46&\cr
&5466&&5.2&&8.91&&281.7&&31.62&&0.8&&0.02&\cr
&5694&&5.4&&0.58&&58&&100&&4.3&&4.87&\cr
&5824&&6.6&&0.52&&260.6&&501.18&&5.3&&43.68&\cr
&5904&&5.6&&0.89&&89&&100&&4.0&&3.74&\cr
&5946&&4.9&&0.23&&45.89&&199.5&&4.8&&6.11&\cr
&6093&&6.0&&0.36&&45.32&&125.89&&5.4&&38.07&\cr
&6121&&4.8&&0.48&&19.11&&39.81&&4.1&&2.54&\cr
&6171&&5.1&&1.0&&31.62&&31.62&&3.5&&1.33&\cr
&6205&&5.8&&1.81&&57.23&&31.62&&3.4&&1.91&\cr
&6218&&5.1&&1.07&&26.87&&25.11&&3.5&&1.42&\cr
&6254&&5.4&&1.07&&26.87&&25.11&&3.8&&2.84&\cr
&6256&&5.3&&0.06&&37.85&&630.95&&6.6&&100.56&\cr
&6266&&5.8&&0.28&&21.73&&77.62&&5.7&&59.07&\cr
&6284&&5.4&&0.25&&49.88&&199.52&&5.2&&16.68&\cr
&6293&&5.6&&0.11&&69.40&&630.95&&6.3&&92.4&\cr
&6325&&5.1&&0.05&&31.50&&630.95&&6.7&&105.4&\cr
&6341&&5.3&&0.51&&40.50&&79.43&&4.4&&5.39&\cr
&6342&&5.1&&0.16&&50.59&&316.22&&5.4&&16.92&\cr
&6366&&4.0&&2.13&&16.91&&7.94&&1.9&&0.07&\cr
&6362&&5.0&&2.95&&37.13&&12.58&&2.2&&0.19&\cr
&6388&&6.2&&0.39&&24.6&&63.09&&5.7&&82.2&\cr
height2pt&\omit&&\omit&&\omit&&\omit&&\omit&&\omit&&\omit&\cr}
\hrule}

\vskip 2 cm
Table 1. Data on globular clusters (continue)
\vskip 0.3cm
\vbox{\offinterlineskip
\hrule
\halign{&\vrule#& \strut\quad\hfil#\quad\cr
height2pt&\omit&&\omit&&\omit&&\omit&&\omit&&\omit&&\omit&\cr
&NGC\hfil& &${\lg\frac {M}{M_{\odot}}}$\hfil& &$r_c$\hfil& &$r_t$\hfil& &$r_t/r_c$\hfil& &$\lg\rho_0 $\hfil& &$\sigma_0$&\cr
&\omit\hfil&&\omit\hfil&& pc\hfil&&pc\hfil&&\omit\hfil&&$M_\odot pc^{-3}$\hfil&&g cm${}^{-2}$&\cr
height2pt&\omit&&\omit&&\omit&&\omit&&\omit&&\omit&&\omit&\cr
\noalign{\hrule}
height2pt&\omit&&\omit&&\omit&&\omit&&\omit&&\omit&&\omit&\cr
&6397&&5.4&&0.03&&18.92&&630.9&&5.3&&2.52&\cr
&6441&&6.2&&0.34&&34&&100&&5.8&&90.3&\cr
&6522&&5.3&&0.10&&39.8&&398.1&&6.1&&53.0&\cr
&6535&&4.2&&0.83&&20.84&&25.11&&3.6&&1.39&\cr
&6541&&5.6&&0.57&&113.7&&199.52&&5.5&&75.88&\cr
&6558&&4.4&&0.08&&25.29&&316.2&&5.6&&13.4&\cr
&6624&&5.2&&0.15&&47.43&&316.2&&5.6&&25.14&\cr
&6626&&5.4&&0.41&&20.54&&50.11&&4.9&&13.71&\cr
&6656&&5.7&&1.23&&24.54&&19.95&&4.0&&5.17&\cr
&6681&&5.2&&0.06&&60&&1000&&6.5&&79.87&\cr
&6712&&5.0&&1.86&&14.77&&7.94&&3.0&&0.78&\cr
&6715&&6.3&&0.66&&66&&100&&5.0&&27.78&\cr
&6752&&5.2&&0.21&&66.4&&316.2&&5.2&&14.01&\cr
&6779&&5.0&&1.04&&26.12&&25.11&&3.4&&1.09&\cr
&6809&&5.3&&3.98&&19.94&&5.01&&2.5&&0.52&\cr
&6838&&4.5&&0.72&&14.36&&19.95&&2.8&&0.19&\cr
&6864&&5.9&&0.51&&51&&100&&4.9&&17.05&\cr
&6934&&5.3&&1.07&&42.59&&39.81&&3.6&&1.79&\cr
&7078&&6.3&&0.21&&132.5&&630.95&&6.2&&140.12&\cr
&7089&&6.0&&1.17&&92.93&&79.43&&4.0&&4.92&\cr
&7099&&5.3&&0.12&&60.14&&501.1&&5.9&&40.12&\cr
height2pt&\omit&&\omit&&\omit&&\omit&&\omit&&\omit&&\omit&\cr}
\hrule}
\vskip 0.7 cm

{\it 2.3. The lens distribution along the line of sight} 
\vskip 0.5cm
Since we are taking globular clusters in galactic halos to be the gravitational lenses, the lens distribution in space is determined by the 
large-scale distribution of galaxies. In recent years, there has been considerable progress in the study of the large-scale structure of the
Universe via massive measurements of redshifts for deep galactic samples (see the reviews [18,19,20]). Statistical analyses of data on the 
three-dimensional distribution of galaxies have shown the presence of fractal structure up to a few hundreds of Mpc [21, 22, 23], and faint
galaxy counts suggest that this fractality continues to distances of 1000 Mpc [20, 24].

The necessity of taking into account the fractal distribution of gravitational lenses in statistical analyses of lensing effects was first
discussed in [25]. Following this work, we choose the following parametric representation for the density of lensing objects along the line 
of sight between the source and the observer: $$n_l(R)=0.5n_{ol}\Bigl[\Bigl(\frac{R}{R_0}\Bigr)^{D_F-3}+
\Bigl(\frac {R_s-R}{R_0}\Bigr)^ {D_F-3}\Bigr],\eqno(5)$$
Here, $n_{ol}(R_0)$ is the density of objects at a distance $R_0,$ $R$ and $R_s$ are the distances to the lens and the source, respectively;
 and $D_F$ is the fractal dimension of the galaxy distribution. The uniform lens distribution considered here is obtained from (5) with
 $D_F=3,$ in which case  $n_l(R)=\hbox{const}.$ The observed value of the fractal dimension is close to $D_F\sim 2$ [19, 20].
\vskip 0.7cm

{\bf 3. LENSING BY KING OBJECTS}
 \vskip 0.5cm
One of the key assumptions in our model is that the gravitational lenses have King mass distributions, described by equations (1). King 
lenses have been analyzed in considerable detail in [17], and we use the results of this work for our lensing probability calculations.
\vskip 0.4cm
{\it 3.1. Basic Equations}
\vskip 0.3 cm
Figure 1 presents a diagram of the general lensing geometry. 

The basic equation of a gravitational lens is the condition for a ray to reach the observer:
$$\alpha(x)=\frac {x-y} {g},\eqno(6)$$
where $x= {l}/ {r_c}$ is the impact parameter in the lens plane in units of the lens core radius $r_c$, $ y= {l_s}/{r_c}$ is  the deviation of the
source $l_s$ from the observer-lens axis in the  lens plane in units of $r_c$;
$\alpha=\delta(l)/\alpha_c$ is the deflection angle $\delta(l)$ of a ray passing   at the impact distance $l$ in units of the angle $\alpha_c=r_c/F_L$ subtended by 
  the lens core at the  lens focus. The lens position parameter $g$ is given by
$$g=\frac{R_{OL}R_{LS}}{R_{OS}F_L},\eqno(7)$$
where $F_L$ is the focal length of the lens and $R_{OL}, R_{LS}$ and $R_{OS}$  are the observer-lens, lens-source, and observer-source distances 
 (the angle $\beta$ is assumed to be small).

For distant objects, we must calculate their distances using some particular cosmological model. We can represent the relation between the angular 
distance $R$ and the redshift $z$ in the general form
$$R=R_H\Psi (z),\eqno(8)$$
where  $\Psi (z)$ is a normalized angular distance that depends on the parameters of the  cosmological model   ($q_0,\Lambda$ и др.),and $R_H=c/H_0$
is the Hubble  distance. There  is not currently a  generally  accepted  opinion  about  the  most appropriate  cosmological  parameters ; 
 for convenience in the  calculations, we adopt here  $\Psi (z)=z,\hskip 10pt H_0=75\hbox{km/s Mpc,}$ and  $R_H=4000\hbox{Mpc.}$
The expression for  the  parameter   $g$  then  takes  the form 
$$g=11.4a(1-a)z_s\Bigl(\frac {\sigma_0}{1[\hbox{\r}]}\Bigr),\eqno(9)$$
where  $a=z_L/z_S$ is the  relative  position of the lens ;  $z_S$ and $z_L$ are the redshifts of the  source and lens, respectively; and $\sigma_0$ is 
the central surface density of the lens given by (2).
   
The gravitational deflection of rays passing at a distance equal to the impact parameter  $l$ is determined by the mass contained in the sphere of radius $l$.
The normalized deflection angle in a King model is

$$\alpha (x)=\frac {\Lambda (x)} {x},\eqno(10) $$
where
$$\Lambda (x)=$$
$$\cases{\ln(1+x^2)-4(1+x_t^2)^{-1/2}\cdot
\Bigl [(1+x^2)^{1/2}-1\Bigr ]+x^2(1+x_t^2)^{-1},&for
$\vert x\vert <x_t$\cr \Lambda (x_t),&for $\vert x\vert \geq x_t$\cr}\eqno(11)$$
Here, $x=l/r_c,\hskip 10pt x_t=r_t/r_c.$

Simultaneous solution of equations (6) and (10) provides information about the number of images and their positions in the lens plane for a fixed parameter $g.$
Figure 2 presents a classification of possible situations for  the case of lensing of a point source by a King lens. 

It is clear from these images that we are 
always capable of seeing one, two, or three source images. The origin of the coordinate system coincides with the lens center, making it possible to determine
the position of the image relative to the lens center. A King model has a focus and two caustics: axial and conic. The point with the coordinates  $ g=1,\ y=0$   
corresponds to the focus of the lens, where all three images merge into a single image (Fig.2, graph ''b'' on the right). All caustics meet at the focus. A single
image forms for various positions of the observer when   $g<1$ , and for  $y>y_k$ when   $g\geq 1$ (region I). For  $g>1$ and  $0<y<y_k$, all three images form
(region \uppercase\expandafter{\romannumeral 2}). When caustics intersect, two images merge into a single image. When $g>1$ and $y=0,$ the  observer is located
at the axial caustic, and we observe a central image and a ring of radius $x$ (Fig. 2, graph  ''o'' on the right).   When  $g>1$ and  $y=y_k$,  the observer is located 
at the conic caustic (Fig.2, graph ''k'' on the right) and there is a merging of two images. When  $y>y_k$, this merged image entirely disappears; in the case $y<y_k$, 
it splits into two separate images.
\vskip 0.7 cm
{\it 3.2. Amplification coefficient}
\vskip 0.5cm
Among the main manifestations of gravitational focusing are included not only image multiplication, but also possible image amplification. The amplification 
coefficient is given by the expression $k=k1\cdot k2,$ where
 $$k1=\d{\frac {dx}{dy}}=\Bigl [1-g\d{\frac {\alpha (x)}{dx}}\Bigr ]^{-1},
 \hskip15pt k2=\d{\frac {x}{y}}=\Bigl [1-g\d{\frac {\alpha (x)}{x}}\Bigr ]^{-1}$$
For $\vert x\vert \ll x_t$ we have, with a high degree of accuracy 
$$k=\Bigl [2-\d{\frac {2g}{1+x^2}}-\d{\frac {y}{x}}\Bigr ]^{-1}\cdot \Bigl
[\d{\frac {y}{x}}\Bigr ]^{-1}\eqno(12)$$

This formula shows that there is infinite amplification at the lens focus 
${(y=0,}\ {g=1),}$  at the axial caustic ${(g>1,}\ {y=0),}$   and at the conic caustic $(\d{\frac{2g}{1+x^2}}+\d{\frac{y}{x}}-2=0)$.
 The infinite amplification is a consequence of our use of the geometric optics approximation. Allowance for any ray divergence leads to a finite amplification. In particular, we can take into account the angular size of the source, 
$\Theta_0$, then 
$$k_{max}\approx\d{ \frac {\xi}{R_{LO}\Theta_0}},\eqno(13)$$
where $\xi=\sqrt{2r_g\d{\frac {R_{OL}R_{LS}}{R_{OS}}}}$ is the radius of the Khvol'son-Einstein ring, and $\Theta_0=r_s/R_{OS}$ is the angular size of a source with linear size $r_s$ located at a distance $R_{OS}.$
It follows from (13) that the extended sources cannot provide higher amplification coefficients, but if we consider the lensed sources to be objects made up of individual stars and interstellar clouds, the total
 amplification will be the sum of the amplification coefficients of the individual components. In this way, both the lines and the continuum in the source spectrum can be amplified by the same factor. If the source size is 
$ 10^{16}\hbox{cm} (\sim 10^{-2}\hbox{ pc})$ and the mass of the lens is  $10^6M_\odot$, then $k_{max}\sim 10^4,$ corresponding to a brightness amplification of $10^m.$
 Figure 3 shows the dependence of the amplification coefficient 
$k$ on the displacement $y$ of the source in the lens plane. 

The amplification coefficient $k$ as a function of the lens position $a$ on the observer-source axis for various values of $y$ 
and $\sigma_0$ is plotted in Fig.4.

We can speak of a significant amplification coefficient only if the observer is not between the focus and the lens, so that $g\geq 1.$ The position of the focus is determined by the central surface density of the lens,
 and thus we can determine the minimum density $\sigma_{0 cr}$ corresponding to the condition 
$g=1.$ Using (4) and (9), we obtain 
$$\sigma_{0cr}(a,z_s)=\frac{0.077}{a(1-a)z_s}\hbox{[\r]}.\eqno(14)$$
According to (14), for any fixed $a$ and $z_s$ of the total distribution $f_{\sigma_0}$ of the lenses in $\sigma_0,$
 only objects with $\sigma_0\geq\sigma_{0cr}$ take part in the lensing.
\vskip 1cm

{\it 3.3. The lens cross-section}
\vskip 0.3cm
A necessary condition for strong gravitational lensing is that the observer be located
 near the axial or conic caustic. The size of this region is determined by the lens
 position parameter $g$ and by the desired value of the amplification coefficient $k.$
 The dimensionless cross section of the caustic regions in the lens plane has two terms:

$$S(k,g)=S_0+S_k=\pi {\Delta y_0}^2+\pi( y_k^2-y_2^2),\eqno(15)$$
Here, $\Delta y_0$ is the radius of the cross section of the axial caustic;
$y_k=y_2(k_{max},g),$\hskip10pt $y_2=y_2(k,g),\hskip 0.3cm S_k$ is the cross section of the conic
 caustic; and $S_0$ is the cross section of the axial caustic. The dimensional cross
 section is obtained by multiplying (15) by ${r_c}^2.$
Figures 5 and 6 present a visual representation of the position of the conic caustic and 
the relation between the conic and axial regions. 

Numerical calculations show that $S_0$
 increases and $S_k$ decreases with increasing $g.$ For a fixed source redshift $z_s$ 
and a central surface density $\sigma_0,$ the value of $g$ increases when $a$ and
 $\sigma_0$ increase. Thus, as the lens moves to the center, the cross section of the 
axial caustic decreases, while the cross section of the conic caustic decreases.
 Similarly, when the lens moves toward the observer or toward the source, the cross
 section of the axial caustic decreases, and the cross section of the conic caustic 
increases. When $g$ grows, the radius of the cross section of the conic caustic also 
grows, while its thickness decreases. Thus, the area of the cross section of the conic 
caustic is determined by the relative rate of these two processes. Since
$\Delta y_k\sim\d{\frac{1}{k^2}},$ we have
$\Delta {y^{\prime}}_k\sim\d{\frac {1}{k^3}}.$ This relationship shows that the rate of change
 of the thickness decreases with increasing $k,$ while the radius of the region $S_k$ is
 determined only by the value of $g$ (Fig. 2), i.e., it does not depend on $k.$ The 
behavior of the total cross section is determined by the behavior of $S_k$ and $S_o.$
 If no constraints are imposed on $y$ the axial caustic dominates over the conic caustic,
 and the function $S(k,g)$ has a maximum at the center (Fig.6). Figure 7 shows the
 dependence of the total cross section on $g$ for various amplification coefficients.

The presence of the conic caustic is advantageous for the King model compared to a point
 lens model (applicable for individual stellar lenses) or an isothermal sphere lens model
 (applicable for spiral galaxy lenses). Because $y_k$ varies between $0$ and $r_t$
 when $a$ is changed from 0 to 1, we can always locate the observer near the conic
 caustic for any given lens position by properly choosing the value of $\sigma_0.$
 Thus, the conic caustic offers greater freedom in moving the source in the lens plane 
and does not require an exact alignment of the observer, source, and lens on the same
 axis. The important role of conic caustics is pointed out in [26].
\vskip 0.7cm

{\it 3.4. Lensing probability}
\vskip 0.3cm
We will now define the number of lenses in the volume 
$dV=4\pi z_l^2dz_l$ with the surface density $\sigma_0$ in the interval
$d\sigma_0$ to be
  $$dN=n_{l} f(\sigma_0) dV d\sigma_0,\eqno(16)$$ 
where $f(\sigma_0)$ is the surface density distribution of the lenses, which, according to
 (5), can be written in the form 
$$n_l=n_{0l}\Bigl [ (z_l)^{D-3}+(z_s-z_l)^{D-3}\Bigr ].\eqno(17)$$
We choose $n_{0l}$ so as to satisfy the relation
$$n_{0l}\cdot \bar M_{GC}=\rho_{cr}\Omega_l,\eqno (18)$$
at $R_0=R_H.$ Here, $\Omega_l$ is the mass fraction that is included in lenses; 
$\rho_{cr}\approx 10^{-7}M_{\odot}\hbox{pc} ^{-3}$ is the critical density of matter; and
 $\bar M_{GC}$ is the mean mass of a globular cluster. Let us assume that all sources are
 located on a sphere of area $4\pi z_s^2.$
We can carry over all cross sections from the lens plane to the source plane; we thus 
obtain the cross section in the source plane,
$$d\hat S=S*\Bigl (\d{\frac {z_s}{z_l}}\Bigr )^2 dN.\eqno(19)$$
Dividing this value by the source sphere area, we obtain the differential probability of 
lensing a source located at the distance $z_s$ by a lens with surface density $\sigma_0$
 at the distance $z_l$, with the amplification coefficient $k,$ and with the fractal 
dimension of the lens distribution along the line of sight, $D_F:$
$$dP(k,z,z_s,D_F)=\frac {d\hat S}{4\pi z_s^2}.\eqno(20)$$
Accordingly, after integrating over all possible lens positions and all allowable 
densities, we obtain the probability of lensing with amplification coefficient exceeding 
$k,$ 

$$P(k,z_s,D_F)=$$
$$=\frac{c}{H_0}n_{0l} r_c^2\int\limits_{\d{\sigma_{0cr}(z_l,z_s)}}
\limits^{\d{\infty}}\int\limits_{\d{z_{*}}}\limits^{\d{z_s-z_{*}}}
S(k,z_l,z_s,\sigma_0) f(\sigma_0)
\Bigl[(z_l)^{D_F-3}+(z_s-z_l)^{D_F-3}\Bigr]d\sigma_0dz_l,\eqno(21)$$

where $z_{*}$ is the initial level of fractality. Let us assume that all the sources have
 $z_s=1$ and the $25\%$ of the mass of matter in the Universe. We will estimate the probability 
of lensing by globular clusters compared to the probability of lensing by ''point''
 objects for the case of a uniform lens distribution. It is clear from Fig.8 that the 
probability of lensing by a King object is comparable to the probability for a point model
 [4]. Figure 8 also shows the dependence of $P(k)$ for King lenses in the case of a 
fractal distribution with dimension two.

 The presence of fractal structure increases the 
probability of lensing by nearly an order of magnitude. This plot also indicates that the
 differential probability function $p(k)=dP/dk$ obeys the law 
$$p(k)dk\sim k^{-3}dk.\eqno(22)$$

Figure 9 shows the dependence of the differential probability (20) on the relative 
position of the lens $a$ on the observer-- source axis for two fractal dimensions, $D_F=2$
 and $D_F=3,$ and two amplifications, $5^m$ and $8^m.$

It is clear that, for a uniform distribution of galaxies in space (curves 1 and 3),
 the most likely position of the lens is halfway between the source and the observer.
 For a fractal distribution with amplification coefficients of the order of $8^m$ and
 higher (curve 4), the location of maximum probability for finding the lens is displaced
 toward the observer and toward the source (the distribution has two humps). At lower
 amplifications for a fractal distribution (curve 2), we obtain a nearly flat distribution.
\vskip 4cm

{\bf 4. INTERPRETATION OF QUASAR-GALAXY ASSOCIATIONS}
\vskip 0.5cm
 Using the catalog of associations [5], we will attempt to obtain some observed properties
 of the associations that must be accounted for by the proposed gravitational lensing 
model. This catalog contains a total of 577 quasars and about 500 galaxies with which the 
quasars form associations. Fourteen quasars have absorption lines with the same $z$ as the
 galaxy. Of these objects, we will select pairs satisfying the requirements:

(1) There are data for $m_Q,\  z_Q,\  z_G$;

(2) $z_G< z_Q$;

(3) The quasars are at angular distances not exceeding the angular distance corresponding 
to a linear distance of 150 kpc.

(4) If a quasar is projected onto several galaxies, we chose the galaxy with the minimum
 linear distance from the quasar.
As many as 241 pairs satisfy these criteria.
Let us first estimate the expected number of associations based on the calculated
 probability of lensing the nuclei of active galaxies. As we noticed in Subsection 2.1, 
the number of active nuclei in galaxies with apparent magnitudes down to $28^m-29^m$ is
 estimated to be $\sim 10^4\hbox{ deg}^{-2}.$
The integrated probability of lensing by globular cluster-type objects with
 amplifications of $5^m-8^m$ is of the order of $10^{-5}-10^{-7}.$ Thus, the expected
 number of quasars in associations could be of the order of 
$\sim 10^2-10^4,$ which is comparable to the observed value ($\sim 500$ pairs.)
This means that quasars with high amplification coefficients are encountered in
 associations, and that quasars in associations may have, on the average, higher
 brightnesses than other quasars.
Figure 10 presents the distribution of the number of quasar-galaxy pairs as a function 
of the parameter $a=z_G/z_Q,$ specifying the relative lens position along the line of 
sight for several fixed halo sizes and several $z_Q$ values.

 We can see a clear tendency
 for associations to concentrate at $a<0.1$ and $a>0.9,$ i.e., there is an excess of 
gravitational lenses near the observer and near the source. This behavior of the 
associations agrees well with the theoretical behavior of the differential lensing
 probability shown in Fig. 9. The smaller number of pairs with $a\sim 0.9$ compared to 
$a\sim 0.1$ may be due to an observational selection effect: the farther away the lens,
 the harder it is to detect it. It is possible that some quasars not included in
 associations are also lensed, since, if a quasar is not in an association, it may be that
 the galaxy is located at a greater distance, nearer to the quasar, and is simply not 
detected. In this case, the galaxy can be revealed from the absorption lines observed in
 the quasar spectra. Figure 11 shows a histogram of the $(1+z_{em})/(1+z_{abs})$ 
 ratio for   quasars with absorptions [27].

 This dependence is in agreement with the
 lensing model, since there is an excess of systems with $z_{em}\sim z_{abs}.$ The 
detection of systems with
$z_{abs}\ll z_{em}$ is difficult, since in this case, the absorption lines are strongly
 shifted toward the ultraviolet. Some observational material is available, however. Among
 14 quasars with absorptions in the Arp catalog of quasars, 7 have $z_{abs}<0.1z_{em}.$

An additional argument in favor of the gravitational lensing model presented here is the
 similarity between the surface density distribution of globular clusters and the critical
 surface density distribution of associations calculated from formula (14). Figure 12 
presents the distributions   
$f_{\sigma 0}$ and $f_{\sigma cr}$  in the range <100 [\r]. 

The critical density is reached 
at the lens focus ($g=1$). Theoretically, such behavior of the predicted critical density
 is expected, since at large amplifications a significant portion of the cross section is
 concentrated in the focal region (Fig. 7).

Thus, we can conclude from the above estimates that quasars in associations may be
gravitationally amplified images of the nuclei of active galaxies, and that globular
 clusters in galactic halos play the role of the gravitational lenses. The arguments 
against gravitational lensing presented in [3] lose their force in the framework of this 
model. First, the quasars are projected onto the galactic haloes because that is where
 globular clusters are primarily located, and the probability of lensing is sufficient for
 reasonable values of $\Omega_L.$ Second, companion galaxies in close galaxy groups may
 have an enhanced probability for the birth of globular cluster-type objects. Third, for
 the catalog of associations [5], there is no relation between the redshifts of the
 quasars and the galaxies. Fourth, since the nuclei of active galaxies are supposed to be 
the sources, there is no problem with the number of lensed objects. Fifth, and finally,
 variability associated with the motion of the globular clusters would be observable only
 on timescales of about $10^3 \hbox{yr}.$ 
\vskip 0.7cm

{\bf 5. DISCUSSION}
\vskip 0.5cm
Since the existing observational data on quasar-galaxy associations are consistent with 
the proposed gravitational lensing model, it is naturally to pose the question of possible 
further observational tests. We can offer the following observational tests for the 
properties of the quasars listed in the catalog [5]:

(1) It follows from the estimated angular radius of the Khvol'son-Einstein ring that
 the expected angular splitting of quasar images by gravitational lensing by globular
 cluster-type objects is of the order of milliarc-seconds:
$$\Theta_{Ch-E}\sim 30 \Bigl(\frac{M}{10^6 M_\odot}\Bigr)^{1/2}
 \Bigl(\frac{D}{10 \hbox{Mpc}}\Bigr)^{-1/2}$$
Here, $D=\d{\frac{R_{OL}R_{OS}}{R_{LS}}}$ is the effective distance to the lens.
For this reason, it is important to examine the structure of compact components on mass
 scales in quasars in associations that have sufficient radio fluxes for VLBI
 observations. Note that the theory does not exclude the occurrence of a single image 
(when the observer is in region I; see Fig. 2). There may also be no observed splitting
 due to inadequate dynamic range of the VLBI images.

(2) The expected flux variability associated with the motion of the globular cluster as a
 whole has a timescale longer than a thousand years, since the estimated characteristic
 variability timescale is given by
\uppercase\expandafter{\romannumeral 1},.2).
$$t_{var}\sim\frac{R_{OL}\Theta_{Ch-E}}{v}\sim 10^3 \Bigl(\frac{M}{10^6M_\odot}\Bigr)^{1/2}
\Bigl(\frac{D^{\prime}}{10 \hbox{[Mpc]}}\Bigr)^{-1/2}
\Bigl(\frac{v}{10^3\hbox{ [km/s]}}\Bigr)^{-1}\hbox{yr},$$
where $D^{\prime}=\d{\frac{R_{LS}R_{OL}}{R_{OS}}}$ is the reduced distance to the lenses and 
$v$ is the velocity of the lens motion. However, there can be flux variability on 
timescales shorter than a year due to microlensing by individual stars in the globular
 clusters. Microlensing effects are more pronounced for quasars associated with nearby 
galaxies, since amplification by point objects depends on the ratio $R_{OS}/R_{OL}$ 
(see formula (13)).

3) Since globular clusters belong to galactic haloes, absorption lines may arise in
 quasar spectra, with the $z_{abs},$ corresponding to the galaxy in which the lensing
 globular clusters are located. It is of interest to continue the work started in [28],
 concerning comparisons of the properties of quasars with absorptions and quasars in
associations. In particular, the known effect that the number of absorption lines
 increases closer to the quasar may result from the fractal character of the large-scale 
distribution of galaxies along the line of sight. In this case, we also would expect an 
increase in the number of absorption lines closer to the observer.

(4) The equal status of the observer and the source inside a fractal structure suggests 
that, in addition to the ''Arp effect'' that distant quasars are observed near nearby 
galaxies, there should be an increased probability of finding near quasars faint galaxies
 with redshifts close to the quasar redshift. This effect may have revealed by Tyson [29],
 but a representative sample of ''Tyson associations'' must be collected and their
 properties compared to the properties of  ''Arp associations.''

(5) Comparing the spectral properties of quasars associations with the properties of the 
active nuclei galaxies of various types may serve as a test for the structure of the
 emission-line formation region, which could be lensed differently by a King object.

(6) Comparing the properties of quasar-galaxy associations in galaxies of various types
 and in galaxy groups with the properties of globular cluster systems in the halos of 
similar galaxies may serve as a test for the reality of globular clusters as the
 gravitation lenses.

(7) Due to refraction of rays in a globular cluster lens, quasars with milliarcsecond 
jet structures, i.e., with jets in the innermost regions of their compact radio sources,
 may display curvature in the jet trajectories that is not associated with the real
 motion of matter ejected from the nucleus.
\vskip 0.8 cm
{\bf 6. CONCLUSION }
\vskip 0.5 cm 
Our calculations for the probability of strong gravitational lensing by globular clusters
 populating the halos of normal galaxies and positioned along the source-observer line of
 sight offer a simple physical interpretation of the Arp effect: the observed occurrence 
rate of quasar-galaxy associations considerably exceeds the expected number of random 
projections. If we interpret the ''Arp associations'' as the effect of mesolensing, the 
quasars observed in the halos of nearby galaxies with a characteristic size of about 
100 kpc are proposed to be images of compact objects in the nuclei of active galaxies 
located at a distance of $z\sim 1,$ which are gravitationally amplified by
$5^m-8^m.$ In this interpretation, the redshifts of quasars have a cosmological origin,
 and there is no need to invoke new physics.

The interpretation of the excess quasar-- galaxy associations as an effect of gravitation
 lensing is possible only if three factors are simultaneously taken into account:

 (1) the use of the cross section of a King gravitational lens (the globular cluster) 
in place of a point or isothermal model; 

(2) use of a fractal lens distribution along 
the line of sight, instead of the standard uniform distribution;

 (3) the number of faint
 galaxies in recent extremely deep counts.

These three fundamental factors are based on modem observational data on globular 
clusters, the three-dimensional galaxy distributions from deep redshift surveys, and
 recent counts of faint galaxies down to $29^m.$ Some consequences of ''hidden''
 (as a consequence of the extremely small angular splitting of the images) 
gravitational mesolensing can be tested with existing observing techniques.
\vskip 1cm
\centerline {REFERENCES}
\vskip 0.4 cm
1. Arp, H., {\it Astrophys. J.,} 1966, vol. 148, p. 321.

2. Arp, H., {\it Quasars, Redshift and Controversies,} Berkeley: Interstellar Media, 1987.

3. Arp, H.,{\it Astron. Astrophys.,} 1990, vol. 229, p. 93.

4. Schneider, P., Ehlers, J., and Faiko, E.E.,{\it Gravitational Lenses,} New York: Springer, 1992.

5. Burbidge, G., Hewitt, A., Narlikar, J.V., and Das Gupta, P., {\it Astrophys. J. Suppl. Ser.,} 1990, vol. 74, p. 675.

6. Canizares, C.R.,{\it Nature,} 1981, vol. 291, p. 620.

7. Vietri, M. and Ostriker, J.R,{\it Astrophys. J.,} 1983, vol. 267, p. 488.

8.  Linder, E.V. and Schneider, P., {\it Astron. Astrophys.,} 1988, 
     vol. 204, p.L8.

9.  Keel, W.C.{\it Astrophys. J.,} 1982, vol. 259, p. LI.

10. Barnothy, J.M., {\it Astron. J.,} 1965, vol. 70, p. 666.

11. Tyson, J. A., {\it Astrophys. J.,}  1981, vol. 248, p. L89.

12. Tyson, J. A., {\it Astron. J.,}  1988, vol. 96, p. 1.

13. Barnothy, J.M., {\it B.A.A.S.,} 1974, vol. 6. p. 212.

14. Baryshev, Yu.V., Raikov, A.A., and Yushchenko, A.V,{\it Gravitational Lenses in the Universe, 31st Liege Int. Astrophys. Coil.,} 1993, p. 307.

15. Djorgovski, S.G. and Meylan, G. {\it Structure and Dynamics of Globular Clusters,} PASP Cont. Ser., 1993, vol. 50.

16. King, I.R., {\it Astron. J.,} 1962, vol. 67, p. 471.

17. Yakovlev, D.G., Mitrofanov, I.G., Levshakov, S.A., and Varshalovich, D.A., {\it Astrophys. Space Sci.,} 1983, vol. 91, p.133.

18. Coleman, P.H. and Pietronero, L.,{\it Phys. Rep.,} 1992, vol. 213, p. 311.

19. Baryshev, Yu.V., Sylos Labini, R, Montuori, M., and Pietronero, L., {\it Vistas Astron.,} 1994, vol. 38, p. 419.

20. Pietronero, L., Montuori, M., and Sylos Labini, R,{\it Proc. Int. Conf.''Critical Dialogues in Cosmology'',} Princeton, June 24-28, 1996 (in press).

21. Di Nella, H., Montuori, M., Paturel, G., Pietronero, L., and Sylos Labini,R, {\it Astron. Astrophys.,} 1996, vol. 308, p. L33.

22. Sylos Labini, R and Amendo, L., {\it Astrophys. J.,} 1996, vol. 468, p.LI.

23. Sylos Labini, R and Pietronero, L., {\it Astrophys. J.,} 1996, vol. 469.

24. Sylos Labini, R, Gabrielli, A., Montuori, M., and Pietronero, L., {\it Physica A} 1996, vol. 226, p. 195.

25. Baryshev, Yu.V., Raikov, A.A., and Tron, A.A., in ,{\it Gravitational Lenses in the Universe, 31st Liege Int. Astrophys. Coil.,} 1993, p. 365.

26. Ingel', L.Kh.,{\it Astron. Zh.,} 1975, vol. 52, p. 727.

27. Junkkarinen, V.,et al., {\it Astrophys. J., Suppl. Ser.,} 1991, vol. 77, p.203.

28. Dravskikh, A.R, {\it Astron. Zh.,} 1996, vol. 73, p. 19.

29. Tyson, J.A. {\it Astron. J.,} 1986, vol. 92, p. 691.

\end{document}